\documentclass[%
 reprint,
 amsmath,amssymb,
 aps,
]{revtex4-2}

\usepackage{graphicx}
\usepackage{dcolumn}
\usepackage{bm}
\usepackage{color}

\begin{document}

\preprint{APS/123-QED}

\title{Conditioned backward and forward times of diffusion with stochastic resetting: a renewal theory approach}

\author{Axel Mas\'o-Puigdellosas}
\affiliation{Grup de Física Estadística. Departament de F\'{\i}sica, Universitat Aut\`{o}noma de Barcelona, 08193 Bellaterra, Spain}

\author{Daniel Campos}
\affiliation{Grup de Física Estadística. Departament de F\'{\i}sica, Universitat Aut\`{o}noma de Barcelona, 08193 Bellaterra, Spain}

\author{Vicen\c c M\'endez}
\affiliation{Grup de Física Estadística. Departament de F\'{\i}sica, Universitat Aut\`{o}noma de Barcelona, 08193 Bellaterra, Spain}

\begin{abstract}
Stochastic resetting can be naturally understood as a renewal process governing the evolution of an underlying stochastic process. In this work, we formally derive well-known results of diffusion with resets from a renewal theory perspective. Parallel to the concepts from renewal theory, we introduce the conditioned backward and forward times for stochastic processes with resetting to be the times since the last and until the next reset, given that the current state of the system is known. We focus on studying diffusion under Markovian and non-Markovian resetting. For these cases, we find the conditioned backward and forward time PDFs, comparing them with numerical simulations of the process. In particular, we find that for power-law reset time PDFs with asymptotic form $\varphi(t)\sim t^{-1-\alpha}$, significant changes in the properties of the conditioned backward and forward times happen at half-integer values of $\alpha$. This is due to the composition between the long-time scaling of diffusion $P(x,t)\sim 1/\sqrt{t}$ and the reset time PDF.
\end{abstract}

\maketitle


\section{Introduction}
The temporal dynamics of point processes are described by successive \emph{events} in the time axis. At each of these events, an action is performed and the system is modified in some way. It may be the flip of a spin in a magnetic system, the breakdown of a machine in a factory or lightning strikes during a storm. 

In many cases, the time spotted between two successive events can be described by independent, identically distributed (iid) random variables. These type of processes are called \emph{renewal process} \cite{Cox62}. The statistics of renewal processes can be analysed in terms of the density function of the time between two events, i.e. the holding time probability density function (PDF). Some examples of these properties are the number of events occurred until time $t$ or the occupation times of different states of the system \cite{GoLu00,WaScDeBa18}. Two important magnitudes in renewal theory are the forward and the backward times, being the time until the next and since the last event, respectively. The statistical properties of the forward and backward time PDFs depend on the properties of the holding time PDF. For instance, if the holding time PDF is exponential, the forward and backward times have been shown to be exponential too in the long $t$ limit \cite{GoLu00}. Fat-tailed holding time PDFs have also been deeply investigated in \cite{GoLu00,WaScDeBa18}.

In the last few years, a particular type of renewal process has been exhaustively studied in the scientific literature: stochastic resetting \cite{MoMaVi17,EvMaSc20}. Resets are usually treated as a renewal process forcing another stochastic process to start again. For instance, a diffusive random walker would follow its path until an event (reset) happens. Then, the walker is set to be at its initial position to diffuse until the next event (reset). Since diffusion with Markovian resets was studied in \cite{EvMa11}, many stochastic processes have been analysed when they are suddenly set to start again. For instance, resetting applied to Levy flights \cite{KuMaSa14,KuGu15,MeCa16}, run-and-tumble \cite{EvMa18,Br20} and sub-diffusive \cite{KuGu19,MaMo19} random walkers have been studied, as well as other types of stochastic processes \cite{MoVi16,MaTh17,Be18,MeBo18,Ma19,MaMaSc20}. Also, different types of resetting mechanisms have been considered, apart from instantaneous Markovian resetting \cite{NaGu16,MaCaMe19,MaCaMe19p,BoChSo19,BoChSo19p,KuTo19}.

The vast majority of these works focus on two measures: the propagator and the mean first passage time of the process. While the first captures the spatial behaviour, the latter provides information about the possible convenience of resetting to facilitate reaching certain interesting areas in a minimum amount of time. Their properties have been extensively studied for different dynamics and resetting mechanisms \cite{EvMaSc20}. While these measures carry important information about the process, renewal theory may provide novel tools to analyse stochastic dynamics with resets. For instance, in \cite{Br20} the occupation time of a run-and-tumble particle with resetting was studied. In this work we introduce the conditioned backward and forward times, being the times since the last and until the upcoming reset, given that the process is currently in a particular state. For a random walker, this would be the backward and forward times conditioned on the walker being at position $x$ and time $t$. This is particularly interesting when it does not depend on the measurement time $t$. In such cases, the properties of the backward and forward times may be induced from the current state only.

With this aim, in the following we consider a random walker described by a propagator $P(x,t)$, starting at $x=0$ which stochastically resets its position at times given by a reset time PDF $\varphi(\tau)$. The resulting motion can be seen as a compound process from two individual stochastic processes: a temporal process for the resetting mechanism and a time-dependent spatial process for the position of the walker between two successive resets. In Section \ref{SecRenewal} we study the temporal dynamics of the process from a renewal theory perspective, and in Section \ref{SecProp} we include the spatial variable. In Section \ref{SecBackFor} we introduce the conditioned backward and forward PDFs and we conclude the article in Section \ref{SecConclusions}.

\section{Renewal Theory for stochastic resetting}
\label{SecRenewal}

A renewal process is a counting process where the times between successive events $\tau_i$, happening at times $t_i$ and $t_{i+1}$ (i.e. $\tau_i=t_{i+1}-t_i$), are variables distributed according to a holding time distribution $\varphi(\tau_i)$. With this simple, general set up one can study multiple features in terms of the distribution $\varphi(\tau_i)$, as the number of events at a given time $t$, or the time of the last and the next event given that the measurement time is $t$.

In this work, we are mainly interested in studying the distribution of the times since the last reset and to the next one at a given time $t$, namely the backward ($B$) and forward ($F$) times, respectively. To do so, we derive an expression for the probability $Q_N(t)$ that the $N$-th event happens at time $t$. For a renewal process, the probability of the $N$-th event at time $t$ is equal to the probability of the $(N-1)$-th event happening at any past time $t-t'$, times the probability of a single event in the remaining time $t'$. This is,
\begin{equation}
Q_N(t)=\int_0^tQ_{N-1}(t-\tau) \varphi(\tau)dt',
\label{re}
\end{equation}
or
\begin{equation}
\hat{Q}_N(s)=\hat{Q}_{N-1}(s) \hat{\varphi}(s),
\end{equation}
where $\hat{f}(s)=\mathcal{L}_s[f(t)]=\int_0^\infty e^{-st}f(t)dt$ is the Laplace transform of $f(t)$ with respect to the variable $t$. This recurrence equation can be easily solved with the initial condition $\hat{Q}_1(s)=\hat{\varphi}(s)$ to obtain
\begin{equation}
\hat{Q}_N(s)=\hat{\varphi}(s)^N.
\label{EqNthReset}
\end{equation}
Summing over $N$ we can get the overall PDF $Q(t)$ of the last event happening at time $t$. In the Laplace space, it reads
\begin{equation}
\hat{Q}(s)=\sum_{N=0}^{\infty} \hat{\varphi}(s)^N=\frac{1}{1-\hat{\varphi}(s)}.
\label{eq:Q-def}
\end{equation}
In renewal theory, $Q(t)$ is called the rate function. In the limit of long times, and assuming the mean holding time is finite, we have that $
Q(t)\sim 1/ \langle \tau \rangle$. This is just the time derivative of the mean number of renewals in the interval $(0,t)$, i.e. $\langle N(t)\rangle \sim t /\langle \tau \rangle$ \cite{WaScDeBa18}.

Now, the probability that exactly $N$ events have happened before the measuring time $t$ and the last reset happened at the previous time $t-B$ is, thus
\begin{eqnarray}
f^B(N,t,B)&=&\int_{0}^{t}dt'Q_{N}(t')\delta(t-t'-B)\int_{B}^{\infty}dy\varphi(y)
\label{EqBackN}
\end{eqnarray}
Now, summing over $N$, the PDF of the backward time $B$ at the measurement time $t$ reads
\begin{eqnarray}
f^B(t,B)&=&\sum_{N=0}^{\infty}\int_{0}^{t}dt'Q_{N}(t')\delta(t-t'-B)\int_{B}^{\infty}\varphi(y)dydt'\nonumber \\
&=&\varphi^*(B)\int_0^t Q(t-t')\delta(t'-B)dt'\nonumber \\
&=&Q(t-B)\varphi^*(B).
\label{EqBack}
\end{eqnarray}
$\varphi^*(t)=\int_t^\infty \varphi(t')dt'$ is the survival probability of the holding time, which in the Laplace space reads
\[
\hat{\varphi}^*(s)=\frac{1-\hat{\varphi}(s)}{s}.
\]
Transforming Eq.\eqref{EqBack} by Laplace in both $t$ and $B$, with conjugate variables $s$ and $u$ respectively, and using Eq.\eqref{EqNthReset} one finds a formal expression for the backward time PDF in terms of the holding time PDF:
\begin{equation}
\hat{f}^B(s,u)=\frac{1}{1-\hat{\varphi}(s)}\frac{1-\hat{\varphi}(s+u)}{s+u}.
\label{EqBackLaplace}
\end{equation}


Let us now derive a similar expression for the forward time distribution $f^{F}(t,F)$, which is the PDF of the time until the next event $F$ given that the measurement time is $t$. In this case,

\begin{equation}
\begin{split}
f^F(N,t,F)=\int_0^tQ_N(\tau)\varphi(t+F-\tau)d\tau.
\end{split}
\label{EqForward0}
\end{equation}
The forward time PDF is then
\begin{eqnarray}
&&f^F(t,F)=\sum_{N=0}^\infty f^F(N,t,F) \nonumber \\
&=&\sum_{N=0}^\infty\int_0^t Q_N(\tau)\varphi(t+F-\tau)d\tau
\label{EqForward}
\end{eqnarray}
which after transforming by Laplace in $t$ and $F$ and using Eq.\eqref{EqBack}, turns into
\begin{equation}
\hat{f}^F(s,u)=\frac{1}{1-\hat{\varphi}(s)}\frac{\hat{\varphi}(u)-\hat{\varphi}(s)}{s-u}.
\label{EqForwardLaplace}
\end{equation}
 Eqs. \eqref{EqBackLaplace} and \eqref{EqForwardLaplace} for the backward and forward time PDFs have been already derived and profoundly studied in \cite{GoLu00,WaScDeBa18}.

\section{Propagator}
\label{SecProp}

The temporal dynamics of a renewal process can be used to describe the resetting of a physical system. Resets are renewals of an underlying stochastic process. Here, we combine the spatial dynamics of a diffusive random walker with the renewal benchmark presented in previous section. For a general, stochastic process with time-dependent PDF $P(x,t)$ and resets happening at random times distributed according to the PDF $\varphi(t)$, the overall propagator can be written as

\begin{eqnarray}
\rho(x,t)&=&\int_0^t f^B(t,B) P(x,B)dB \nonumber\\
&=&\int_0^t Q(t-B)\varphi^*(B) P(x,B)dB,
\label{eq:generalpropagator-def}
\end{eqnarray}
where $f^B(t,B)$ is the distribution of the last reset (event) happening at time $t-B$, i.e. the backward time PDF as defined in Section \ref{SecRenewal}.

The propagator of a diffusive random walker is
\begin{equation}
P(x,t)=\frac{1}{\sqrt{4\pi D t}}e^{-\frac{x^2}{4Dt}}
\label{eq:diff_prop}
\end{equation}
so that, its Fourier transform for the spatial variable reads
\begin{equation}
\tilde{P}(k,t)=\int_{-\infty}^{\infty}  e^{-ikx} P(x,t)= e^{-k^2Dt}.
\end{equation}
Introducing this expression into the Fourier transform of Eq.\eqref{eq:generalpropagator-def} we get 
\begin{equation}
    \tilde{\rho}(k,t)=\int_0^t Q(t-B)\varphi^*(B)e^{-k^2DB}dB.
    \label{ftro}
\end{equation}
And performing the Laplace transform on the time variable, one gets the global propagator in the Fourier-Laplace space in terms of the reset time PDF:
\begin{equation}
    \hat{\tilde{\rho}}(k,s)=\frac{1-\hat{\varphi}(s+Dk^2)}{s+Dk^2}\frac{1}{1-\hat{\varphi}(s)}=\frac{\hat{\varphi}^*(s+Dk^2)}{1-\hat{\varphi}(s)},
    \label{eq:generalpropagator}
\end{equation}
where we have used Eq.\eqref{eq:Q-def} to express the result in terms of the backward time PDF. Then, the behaviour of $\rho(x,t)$ exclusively depends on the reset time PDF. 

In the following we study the overall propagator for different types of reset time PDFs. We first analyse distributions with finite first moment to later consider distributions with all the moments diverging.

\subsection{$\varphi(t)$ with finite first moment}

\begin{figure}
    \centering
    \includegraphics[scale=0.6]{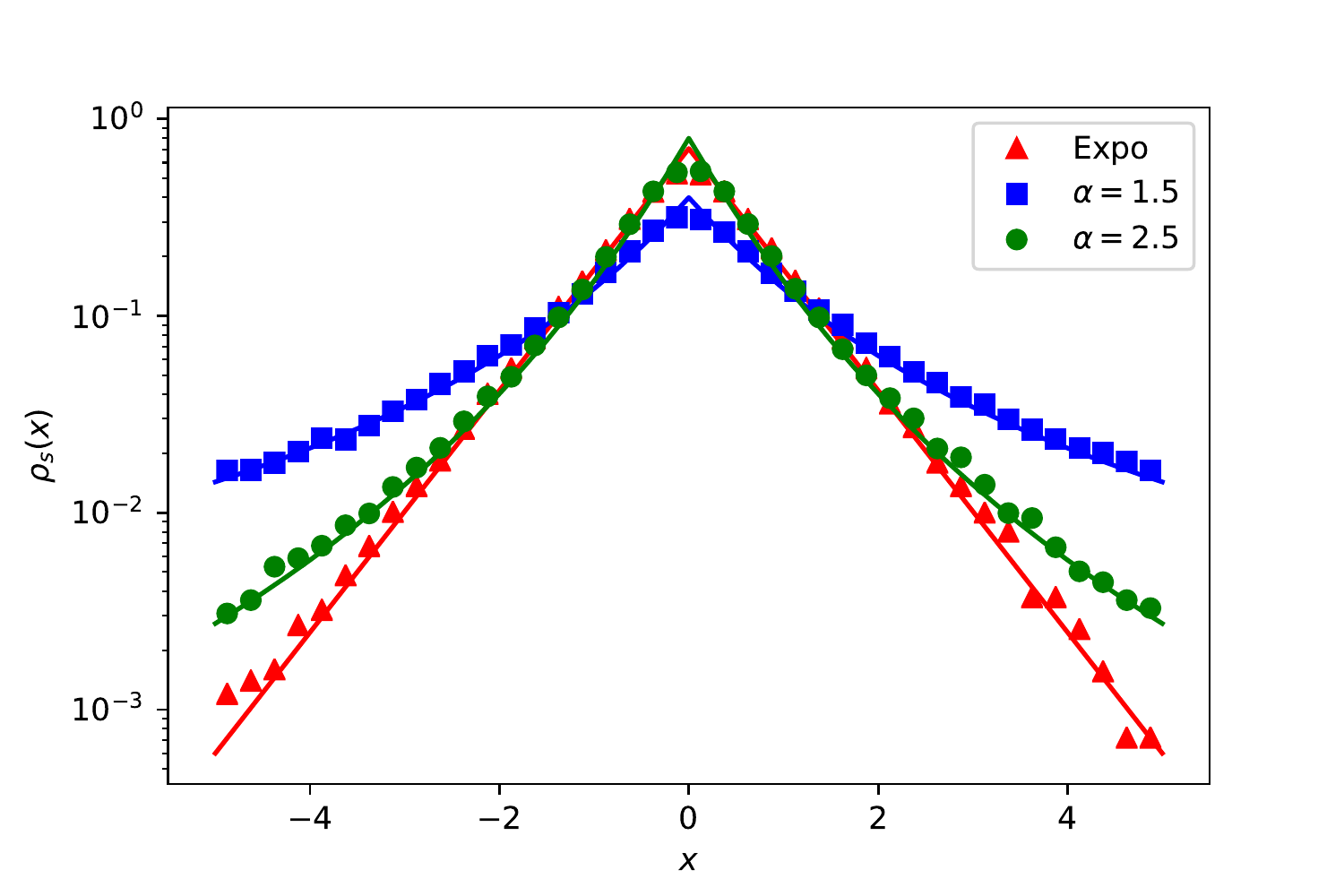}
    \caption{Log-Lin plot of the stationary distribution for diffusion with an exponential reset time distribution (red) and two Pareto reset time distributions with different decay exponents $\alpha$ (green and blue). All the simulations have been performed with $D=0.5$, $T=1$ ($r=1$ for the exponential) and measurement time $t=10^4$. The solid curves show the analytical results in Eqs. \eqref{eq:ststate_expo} and \eqref{eq:ststate_powerlaw}.}
    \label{fig:ststate}
\end{figure}

  We start by considering reset time PDFs with finite first moment.  This includes the cases where all the moments are finite (e.g. Markovian resetting $\varphi(t)=r e^{-rt}$), but also the cases where only the first moment converges (e.g. a power-law PDF decaying as $\varphi(t)\sim t^{-1-\alpha}$ for large $t$ with $1<\alpha<2$). In all these cases, the reset time PDF can be expanded in the Laplace space as $\hat{\varphi}(s)\approx 1-\langle t \rangle_\varphi s +o(s)$. In the long time limit (small $s$), Eq.\eqref{eq:generalpropagator} reads
\begin{equation}
    \hat{\tilde{\rho}}(k,s)\simeq  \frac{\hat{\varphi}^*(Dk^2)}{\langle t \rangle_{\varphi} s}
    \label{12}
\end{equation}
Taking into account that
\begin{eqnarray}
\hat{\varphi}^*(Dk^2)=\int_0^\infty e^{-Dk^2t}\varphi^*(t)dt,
\label{deffe}
\end{eqnarray}
and applying the inverse Fourier and Laplace transforms for the spatial and time variables respectively, the propagator reads
\begin{align}
    \rho(x,t)&\simeq \frac{1}{2\pi \langle t \rangle_\varphi}\int_{-\infty}^{\infty}dk e^{ikx}\int_{0}^{\infty}dt' e^{-Dk^2t'} \varphi^*(t')\nonumber \\
    &=\frac{1}{\langle t\rangle_\varphi}\int_0^{\infty} dt' \frac{\varphi^*(t') }{\sqrt{4\pi Dt'}}e^{-\frac{x^2}{4Dt'}}.
\end{align}
Since this is time independent, a stationary state is reached in this scenario, and the propagator is given by 
\begin{equation}
    \rho_{s}(x) =\frac{1}{\langle t\rangle_\varphi}\int_0^{\infty} dt'  \frac{\varphi^*(t') }{\sqrt{4\pi Dt'}}e^{-\frac{x^2}{4Dt'}}.
    \label{eq:ststate}
\end{equation}
This is general for any type of reset time PDF with finite first moment. Therefore, under this condition, the resetting is always capable of stopping the expansion of the diffusion and reach a stationary distribution in space. Let us consider specific types of resetting PDFs. To deal with Markovian resetting we consider the exponential PDF, i.e. 

\begin{equation}
    \varphi(t)=re^{-rt},
    \label{eq:exponential}
\end{equation}
where $r$ is the constant reset rate. Inserting Eq.\eqref{eq:exponential} into Eq.\eqref{eq:ststate} one has 
\begin{equation}
    \rho_{s}(x)=\frac{1}{2}\sqrt{\frac{r}{D}}  e^{-|x|\sqrt{r/D}},
    \label{eq:ststate_expo}
\end{equation}
which is a well-known result in the resetting literature \cite{EvMa11}. Otherwise, for non-Markovian resetting we consider a power-law reset time PDF of the form 
\begin{eqnarray}
\varphi(t)=\frac{\alpha/T}{(1+t/T)^{1+\alpha}}
\label{eq:pareto}
\end{eqnarray}
with $\alpha >1$ (i.e. a Pareto type  II or  Lomax distribution), such that
\begin{equation}
    \varphi^*(t)=\frac{1}{(1+t/T)^{\alpha}},
    \label{eq:pareto_survival}
\end{equation}
then
\begin{equation}
    \rho_{s}(x)=\frac{\alpha-1}{\sqrt{4\pi DT }} \Gamma\left(\alpha - \frac{1}{2}\right) U\left(\alpha - \frac{1}{2},\frac{1}{2},\frac{x^2}{4DT}\right).
    \label{eq:ststate_powerlaw}
    \end{equation}
Here, $U(a,b,z)$ is the Tricomi's (confluent hypergeometric) function (see chapter 13 in Ref. \cite{ab64}). This result was also found in \cite{NaGu16} by similar means. In the limit $x\rightarrow 0$ the argument of the Tricomi function is small and using the expansions in section 13.1 of Ref. \cite{ab64} we have
\begin{eqnarray}
\rho_{s}(x)\simeq\frac{(\alpha-1)\Gamma(\alpha-1/2)}{\sqrt{4DT}\Gamma(\alpha)}-\frac{\alpha-1}{2DT}|x|
\label{rss}
\end{eqnarray}
and at the resetting point it reaches a finite value
\begin{eqnarray}
\rho_{s}(0)=\frac{(\alpha-1)\Gamma(\alpha-1/2)}{\sqrt{4DT}\Gamma(\alpha)}.
\end{eqnarray}
When the argument is large, i.e., $x^2\gg 4DT$, using Eq. 13.1.8 of Ref. \cite{ab64}
we obtain
\begin{eqnarray}
\rho_{s}(x)\simeq\frac{(\alpha-1)\Gamma(\alpha-1/2)}{\sqrt{4\pi DT}}\left(\frac{\sqrt{4DT}}{|x|}\right)^{2\alpha-1}.
\label{rsl}
\end{eqnarray}
Eqs. \eqref{eq:ststate_expo} and \eqref{eq:ststate_powerlaw} are in agreement with numerical simulations of the process as shown in Fig.\ref{fig:ststate}. The stationary state in Eq. \eqref{eq:ststate_powerlaw} is reached if $\alpha>1$ and hence the solution is time independent. However, when $\alpha <1$ a natural steady state does not always exist, as we show below.

\subsection{$\varphi(t)$ with all diverging moments }
We consider now the scenario where all the moments of the reset time diverge. We can express the distribution in the Laplace space as
\begin{equation}
\hat{\varphi}(s)\approx 1- b_\alpha s^\alpha \quad \text{with}\quad 0<\alpha<1,
\label{16}
\end{equation}
for small $s$. This corresponds to a power law decay of the form $t^{-1-\alpha}$ as $t\rightarrow \infty$. For a Pareto distribution, $b_\alpha=T^\alpha \Gamma(1-\alpha)$ with $\alpha <1$. Plugging this result into Eq.\eqref{eq:Q-def} we have that
\begin{equation}
\hat{Q}(s)\simeq \frac{1}{\Gamma(1-\alpha)(sT)^{\alpha}},\quad sT \ll 1 .
\label{eq:Q-diverging-Laplace}
\end{equation}
Applying the inverse Laplace transform we get
\begin{equation}
Q(t)\simeq\frac{T^{-\alpha}}{\Gamma(1-\alpha)\Gamma(\alpha)t^{1-\alpha}},\quad t\gg T,
\label{eq:Q-diverging}
\end{equation}
which is decreasing with measurement time. Let us analyze  $\rho(x,t)$ by studying the bulk of the distribution and its tail separately. 

\subsubsection{Bulk: $t \gg x^2/D$ and $t\gg T$}

When the measurement time is much larger than the position, the integral in Eq.\eqref{eq:generalpropagator-def} can be simplified with Eq.\eqref{eq:Q-diverging}. Working out the resulting integral one gets to the following expression for the propagator in this region
\begin{eqnarray}
\rho(x,t)&\simeq&\frac{1}{\sqrt{4\pi D}}\frac{1}{\Gamma(1-\alpha)\Gamma(\alpha)}\nonumber\\
&\times&\int_{0}^{t}\frac{e^{-\frac{x^{2}}{4DB}}}{\sqrt{B}\left(t-B\right)^{1-\alpha}\left(T+B\right)^{\alpha}}dB.
\label{eq:prop_bulkdiverging}
\end{eqnarray}
 If we analyse the long time limit $t\gg T$, the propagator in the bulk region can be expressed as (see Appendix \ref{AppendixA}.1 for further details)

\begin{equation}
    \rho(x,t)\simeq \left\{ \begin{matrix}
\frac{\Gamma\left(\frac{1}{2}-\alpha\right)}{2\pi\Gamma(1-\alpha)}\frac{1}{\sqrt{Dt}}\quad\textrm{if}\quad0<\alpha<\frac{1}{2} \\
 \\
\frac{\sin(\pi\alpha)\Gamma(2\alpha-1)}{\pi\Gamma(\alpha)(Dt)^{1-\alpha}}\frac{1}{|x|^{2\alpha-1}},\quad\textrm{if}\quad\frac{1}{2}<\alpha<1.
\end{matrix} \right.
\label{eq:prop_infmoments_asymp}
\end{equation}

It is worth noting that the long time behaviour of the propagator does not have a unique expression for all values of $\alpha <1$. While for $\alpha > 1/2$ its decays as $t^{-1+\alpha}$, when $\alpha < 1/2$ it does as $t^{-1/2}$, i.e., independent of the exponent $\alpha$. This change of behaviour below or above $\alpha = 1/2$ have been observed to appear for various measures of a diffusive process under resetting times drawn from the Pareto PDF given in Eq. \eqref{eq:pareto}. For instance, in \cite{MaCaMe19}, the mean first passage time is shown to converge only when $\alpha > 1/2$ and the PDF at the origin $\rho(0,t)$ have been seen to behave differently for $\alpha < 1/2$ and $\alpha > 1/2$ \cite{BoChSo19}. The two distinct behaviours of Eq.\eqref{eq:prop_infmoments_asymp} have been numerically studied and the results are presented in Fig.\ref{fig:paretopropagator_tail}. There, it is shown that for $\alpha < 1/2$ the propagator tends to be flat when $x^2\ll 4Dt$, while it behaves as $\rho(x,t)\sim 1/|x|^{2\alpha -1}$ when $\alpha > 1/2$.

\subsubsection{Tail: $x^2 \propto 4Dt$ and $t\gg T$}

When $x^2$ is comparable to $D t$, one has to deal with Eq.\eqref{eq:generalpropagator} differently. In this scenario both $t$ and $x^2/D$ are large, so $s$ and $s+Dk^2$ are small enough to consider that
\begin{equation}\hat{\varphi}(s+Dk^2)\simeq 1-b_\alpha (s+Dk^2)^\alpha.
\end{equation}
Then, the propagator in the Fourier-Laplace space can be approximated by
\begin{equation}
\hat{\tilde{\rho}}(k,s)\approx \frac{1}{s^\alpha (s+Dk^2)^\alpha}=\frac{1}{s}g\left( \frac{k}{\sqrt{s}}\right),
\label{eq:propscaling}
\end{equation}
where we have introduced the scaling function
\begin{equation}
g(\chi)=\frac{1}{(1+D \chi^2)^{1-\alpha}}.
\label{eq:gdeff}
\end{equation}

\begin{figure*}
    \centering
    \includegraphics[scale=0.57]{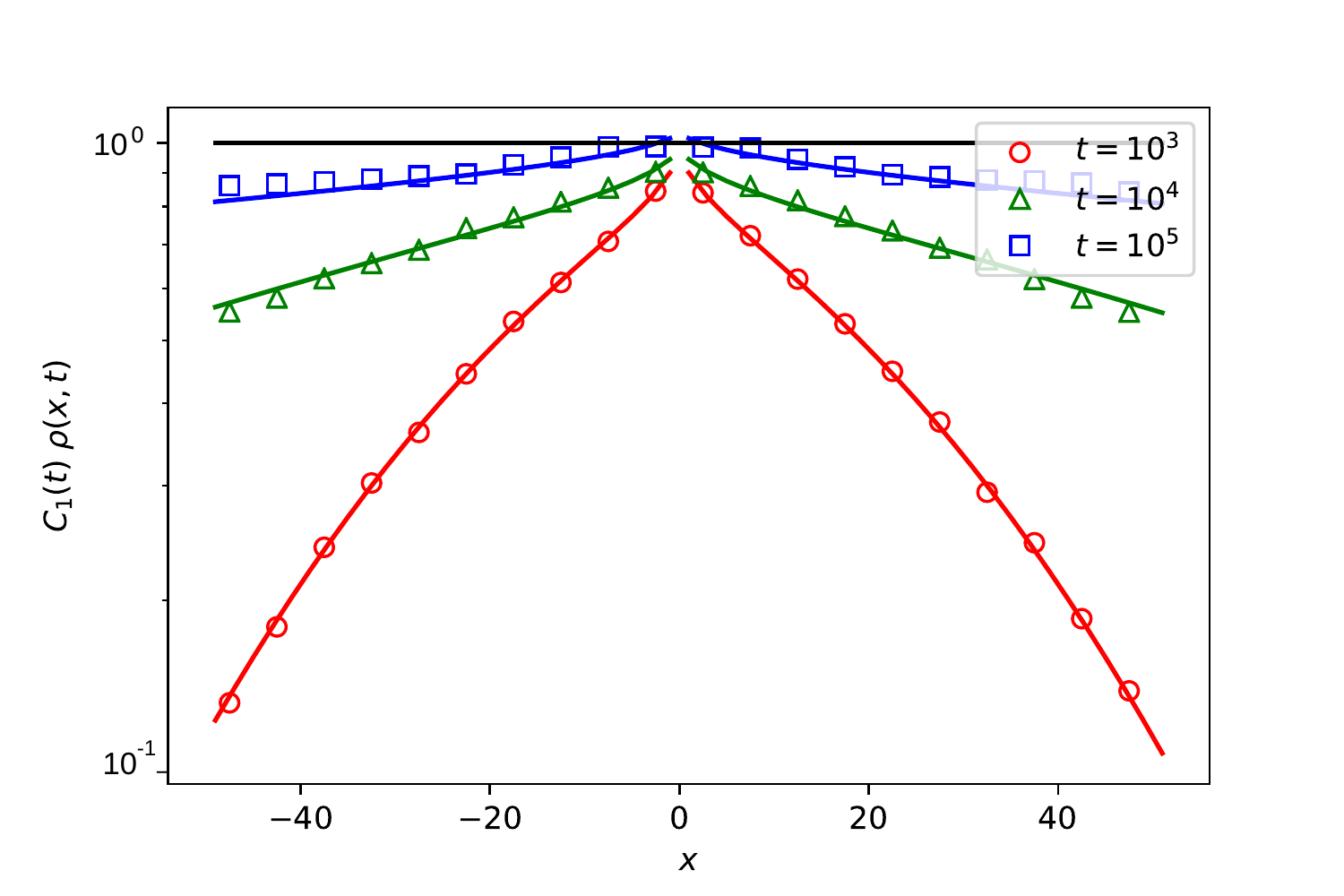}
    \includegraphics[scale=0.57]{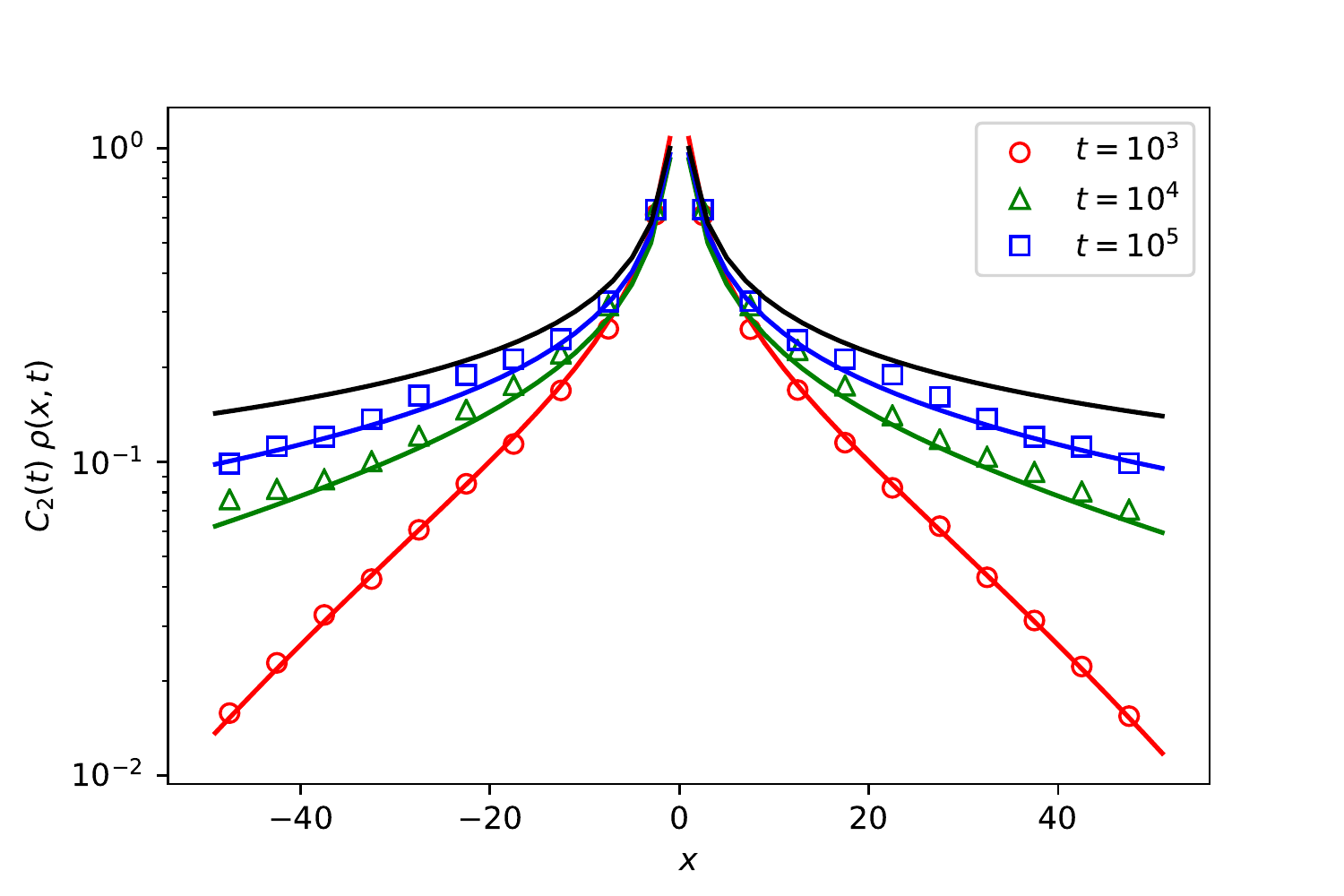}
    \caption{\textbf{A}: Propagator of diffusion with Pareto resetting ($\alpha =0.25$) for three simulations with different measurement times $t$. The multiplicative factor in the y-axis is $C_1(t)=\frac{2\pi \Gamma(1-\alpha)}{\Gamma(1/2-\alpha)}\sqrt{Dt}$. As the time increases, the renormalized propagator tends to be flat as found in Eq.\eqref{eq:prop_infmoments_asymp}. \textbf{B}: Propagator of diffusion with Pareto resetting ($\alpha =0.75$) for three simulations with different measurement times $t$. The multiplicative factor in the y-axis is $C_2(t)=\frac{\pi \Gamma(\alpha)(D t)^{1-\alpha}}{\sin(\pi \alpha)\Gamma(2\alpha -1)}$. As time increases, the propagator approaches this scaling limit. In both panel A and panel B, the diffusion constant is $D=0.5$ and $T=1$ in the Pareto reset distribution. The solid colored lines have been drawn from Eq.\eqref{eq:finalprop_powerlaw}, while the dots, the triangles and the squares correspond to the propagator obtained from the simulations. The solid black curves correspond to the asymptotic behaviour found in Eq.\eqref{eq:prop_infmoments_asymp}. Both panels have been plotted in Log-Lin axis.}
    \label{fig:paretopropagator_tail}
\end{figure*}

In Appendix \ref{AppendixA}.2, we analyse this expression following the same procedure as in \cite{GoLu00}. Doing so, we get the formula for the propagator when $x^2\propto 4Dt$ to be

\begin{equation}
    \rho(x,t)\simeq\frac{1}{\Gamma(1-\alpha)}\frac{e^{-\frac{x^{2}}{4Dt}}}{\sqrt{4\pi Dt}}U\left(\alpha,\frac{1}{2}+\alpha,\frac{x^{2}}{4Dt}\right).
    \label{eq:finalprop_powerlaw}
\end{equation}
This result was found in \cite{NaGu16} by different means. This has been compared to numerical simulations of the process for two distinct exponents $\alpha$. The results are shown in Fig. \ref{fig:paretopropagator_tail}.

\section{Conditioned backward and forward time}
\label{SecBackFor}

Renewal theory provides information about the backward and forward times of a temporal process given that the current absolute time (measurement time) since the dynamics started is known. Nevertheless, in some scenarios, the measurement time may not be an available quantity. For instance, for a random walker which occasionally resets its position by returning to the origin, it is clear how to determine the current position, while it may not obvious how to determine the time since its motion started. With this in mind, in this section we introduce the concepts of \emph{conditioned forward and backward time PDFs}, being the PDF for the forward and backward times given that the current state of the system $x$ is known. We study this magnitude for different types of reset time PDF and determine the conditions under which a stationary form for long $t$ is attained. This is particularly relevant since it permits the study of the backward and forward times by only knowing the current state, independently of the measurement time of the process.

\subsection{Conditioned backward time}

\label{SecBackward}
Let us start by computing the time since the previous reset, given that the walker is currently at position $\mathbf{X}(t)$. This is,
\begin{equation}
    p(B|x,t)=\frac{p(B,x|t)}{p(x|t)}=\frac{p(x|B,t)p(B|t)}{p(x|t)},
    \label{eq:BackBayes}
\end{equation}
where Bayes' law has been employed twice. Now, we define $f(B|x,t)=p(B|x,t)$ to be the conditioned backward time PDF. In the right hand side, $p(B|t)=f^B(t,B)$ is the backward time PDF, $p(x|t)=\rho(x,t)$ is the propagator of the process with resets and $p(x|B,t)=P(x,B)$ is the gaussian propagator. Note that the latter is independent of $t$ since the position of the walker only depends on the time elapsed since the last reset. Introducing the expression for the backard time PDF in Eq.\eqref{EqBack}, one gets

\begin{equation}
    f(B|x,t)=\frac{p^B(B,x|t)}{\rho(x,t)}= \frac{Q(t-B)\varphi^*(B)P(x,B)}{\rho(x,t)}.
    \label{eq:condback_general0}
\end{equation}

In the following, we derive $f(B|x,t)$ from a different perspective, which will be used afterwards to obtain the equivalent distribution for the conditioned forward time. We start by computing $p(B,x|t)$ in Eq.\eqref{eq:BackBayes}, this is the joint PDF of the walker being at position $x$ and that the last reset happened at a previous time $t-B$, given that measurement time is $t$. As for the usual backward time, we start by finding the equation when exactly $N$ resets have happened before the measuring time, i.e.

\begin{eqnarray}
& &p^B(N,B,x|t)=\nonumber\\
& &\int_0^tQ_N(\tau)\delta(t-\tau-B)\varphi^*(B)P(x,B)d\tau,
\label{eq:BackNx}
\end{eqnarray}
where we have included the condition that the walker must be at position $\mathbf{X}(B)$ at the measurement time $t$. Adding all the contributions for $N$ we get
\begin{eqnarray}
& &p^B(B,x|t)=\nonumber\\
& &\sum_{N=0}^\infty\int_0^tQ_N(\tau)\delta(t-\tau-B)\varphi^*(B)P(x,B)d\tau.
\label{EqBackx}
\end{eqnarray}
Applying the Laplace transform and using Eq.\eqref{EqNthReset} one obtains
\begin{equation}
\begin{split}
\hat{p}^B(B,x|s)&=\varphi^*(B)P(x,B)e^{-s B}\sum_{N=0}^\infty \hat{Q}_N(s)\\
&=\frac{e^{-sB}}{1-\hat{\varphi}(s)}\varphi^*(B)P(x,B)
\end{split}
\label{eq:BackxLap}
\end{equation}
and applying the inverse Laplace transform to the result back to $t$, taking Eq.\eqref{eq:Q-def} into account, we readily obtain
\begin{equation}
p^B(B,x|t)=Q(t-B)\varphi^*(B)P(x,B)
\label{eq:BackxLap0}
\end{equation}

It is easy to see that integrating Eq. \eqref{eq:BackxLap0} over $x$ one obtains the marginal PDF for $B$ given in Eq.\eqref{EqBack}. On the other hand, the marginal PDF obtained by integrating $\hat{p}^B(B,x|s)$ over $B$ is the propagator of a process with resets given by Eq. \eqref{eq:generalpropagator-def}. Now, the backward time distribution conditioned on the walker being at position $x$ at time $t$ reads

\begin{equation}
    f(B|x,t)=\frac{p^B(B,x|t)}{\rho(x,t)}= \frac{Q(t-B)\varphi^*(B)P(x,B)}{\rho(x,t)},
    \label{eq:condback_general}
\end{equation}
which is the same as Eq.\eqref{eq:condback_general0}.

Let us study the long time limit ($t\gg B$) of this expression, where we have that $Q(t-B)\simeq Q(t)$. This approximation is accurate for systems where the measurement time scale of the process is many orders of magnitude smaller than the time elapsed since it started. Thus, in this scenario,

\begin{equation}
    f(B|x,t)\simeq \frac{Q(t)}{\rho(x,t)}\varphi^*(B)P(x,B).
    \label{eq:condback_generalbis}
\end{equation}
The time dependence of $f(B|x,t)$ comes from the normalization factor only. So, if $Q(t)/\rho(x,t)$ attains a stationary value for long $t$, then a stationary conditioned backward time PDF $f(B|x)$ exists which only depends on the current position $x$. In the following we analyse the characteristics of the conditioned backward time PDF for different reset time distributions.

\subsubsection{$\varphi(t)$ with finite first moment}

When the reset time PDF has a finite first moment, the diffusion process attains the stationary state given by Eq.\eqref{eq:ststate}. Also, in the asymptotic limit  $t\gg T$, one has $ Q(t)\simeq 1/\langle t\rangle_\varphi$. Therefore, the ratio $Q(t)/\rho(x,t)$ reaches a stationary value and from Eq.\eqref{eq:condback_generalbis}, the conditioned backward time PDF tends to
\begin{equation}
f(B|x)\simeq\frac{\frac{\varphi^*(B)}{\sqrt{B}}e^{-\frac{x^2}{4DB}}}{\int_{0}^{\infty} dt' \frac{\varphi^*(t')}{\sqrt{t'}}e^{-\frac{x^2}{4Dt'}}},
\quad \quad t\gg T, B
    \label{eq:backward_finmoment}
\end{equation}
considering a diffusive process (Eq.\eqref{eq:diff_prop}).

\begin{figure*}
    \centering
    \includegraphics[scale=0.5]{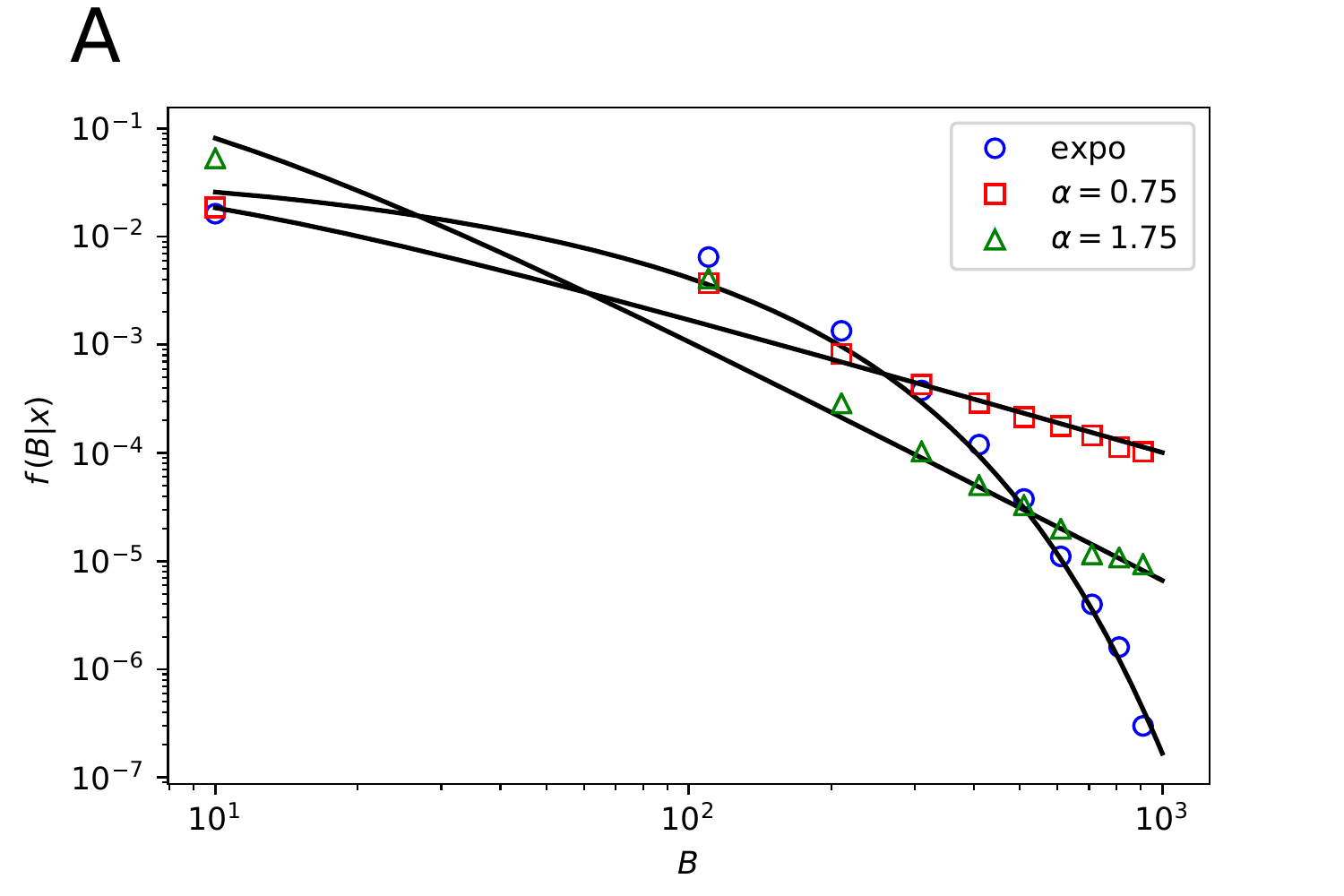}
    \includegraphics[scale=0.5]{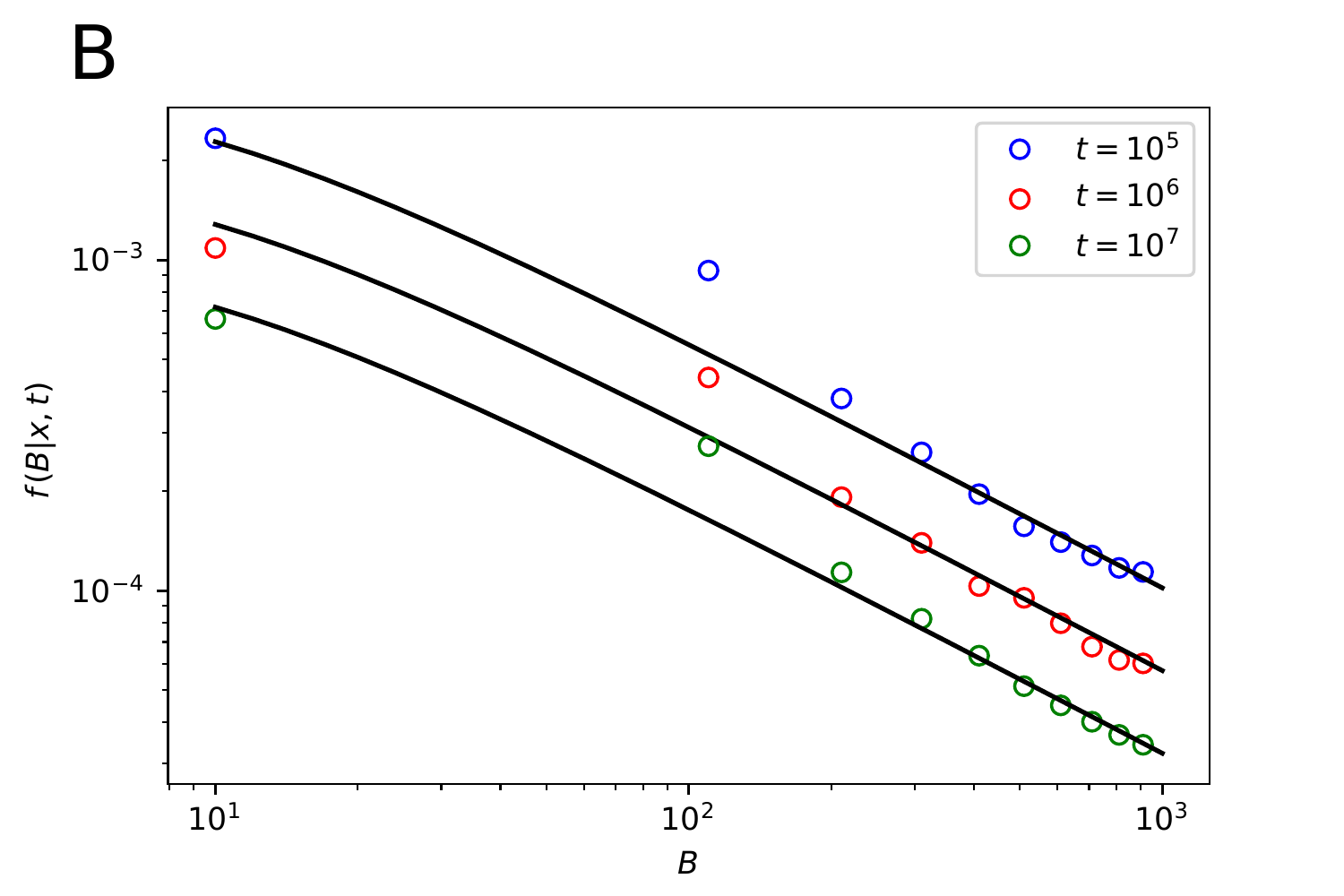}
    \caption{\textbf{A}: Stationary conditioned backward time distribution for diffusion and exponential resetting with $ r=0.05$ and Pareto resetting with $ T=5$ and two different values of the decay exponent $\alpha$. The solid lines have been drawn from the behaviours in Eq.\eqref{eq:statbackward_expo} and Eq.\eqref{eq:statbackward_pareto} for the exponential and Pareto cases respectively. 
    \textbf{B}: Conditioned backward time PDF for diffusion and Pareto resetting with $ T=5$ and $\alpha=0.25$. Three different measurement times have been plotted, showing that $f(B|x,t)$ does not reach a stationary shape. The solid lines correspond to the behaviour in Eq.\eqref{eq:statbackward_pareto} with the time-dependent normalization defined in Eq.\eqref{eq:backwardnorm_pareto}. Both the panels A and B have been simulated with diffusion constant $D=0.1$.} 
    \label{fig:condbackward}
\end{figure*}

When the resetting PDF is exponential, the stationary PDF behaves as

\begin{equation}
    f(B|x)\sim\frac{e^{-rB-\frac{x^2}{4DB}}}{\sqrt{ B}},
    \quad \quad t\gg T, B.
    \label{eq:statbackward_expo}
\end{equation}
So, the conditioned backward time is no longer a Markovian random variable. This is due to the information provided by the position of the walker, similarly to what is found for the unconditioned backward time PDF \cite{WaScDeBa18}. In Fig. \ref{fig:condbackward}A we compare Eq.\eqref{eq:statbackward_expo} (solid line) with numerical simulations (circles). An excellent agreement is observed.

If the reset time PDF is a Pareto distribution with $\alpha > 1$, the behaviour of the stationary PDF with $B$ is as follows:

\begin{equation}
    f(B|x)\sim \frac{e^{-\frac{x^2}{4DB}}}{\sqrt{B}(1+B/ T)^\alpha},
      \quad \quad t\gg T, B,
    \label{eq:statbackward_pareto}
\end{equation}
which has been corroborated by numerical simulations of the process in Fig. \ref{fig:condbackward}A (triangles). We can identify two different regimes. For small $B$, such that $B\propto x^2/4D$, the diffusion process dominates the behaviour of $f(B|x)$ and its shape is Gaussian. However, when $B\gg x^2/4D$, the effect of the resetting becomes more important and the resulting conditioned backward time PDF is a power law of the form 

\begin{equation}
    f(B|x)\sim \frac{1}{B^{\alpha +1/2}}\quad \quad \textrm{as}\quad B\gg x^2/4D .
\end{equation}
Note that when $1<\alpha<3/2$, while the reset time PDF has a finite mean, the conditional mean $\left\langle B|x\right\rangle$ apparently diverges. Nevertheless, $f(B|x)$ is the asymptotic distribution of $f(B|x,t)$, which has a cut-off at $t$. Therefore, even when $t$ is large, the cut-off prevents the first moment of the conditioned backward time PDF to diverge.

\subsubsection{$\varphi(t)$ with all diverging moments}
Let us study the behaviour of the conditioned backward time PDF for power-law resets of the form Eq.\eqref{eq:pareto_survival} with infinite first moment ($\alpha <1$). In this case, if $t\gg x^2/D B$, we have
\begin{equation}
    \frac{Q(t)}{\rho(x,t)}\simeq \left\{ \begin{matrix}\frac{2\pi}{\Gamma(\alpha) \Gamma\left(\frac{1}{2}-\alpha\right)}\frac{\sqrt{D}}{ T^\alpha t^{\frac{1}{2}-\alpha}}\quad\textrm{if}\quad0<\alpha<\frac{1}{2} \\
 \\
\frac{\Gamma(\alpha)}{\Gamma(2\alpha-1)}\frac{|x|^{2\alpha-1}}{D^{1-\alpha} T^\alpha},\quad\textrm{if}\quad\frac{1}{2}<\alpha<1.
\end{matrix} \right.
    \label{eq:backwardnorm_pareto}
\end{equation}

So, for $\alpha \leq 1/2$ , it decays as $t^{\alpha-1/2}$ and consequently $f(B|x,t)$ does not reach a stationary distribution in this case. In Fig. \ref{fig:condbackward}B we show numerical simulations on which the conditioned backward time PDF varies with $t$ even in the long $t$ limit. However, it does reach a stationary distribution when $1/2<\alpha <1$. This case is shown in Fig. 3A where we compare the result in Eq. \eqref{eq:statbackward_pareto} with the numerical simulations for $\alpha=0.75$ (squares). Here, the ratio $Q(t)/\rho(x,t)$ is finite when $t\rightarrow \infty$ and therefore $f(B|x,t)$ converges to the distribution in Eq.\eqref{eq:statbackward_pareto} as when $\alpha > 1$ (see Fig. \ref{fig:condbackward}B).

      

\subsection{Conditioned forward time}
\label{SecForward}

We can proceed similarly for the forward time PDF. In this case, we are interested in knowing the time $F$ until the upcoming reset, given that we know the position of the walker and the measurement time $t$. We start again by computing the joint PDF of the walker being at position $x$ at time $t$ and the following reset happening at time $t+F$. Given that exactly $N$ resets have occurred since the process started, the joint PDF reads:

\begin{equation}
p^F(N,F,x|t)=\int_0^tQ_N(t')\varphi(t+F-t')P(x,t-t')dt',
\end{equation}
which is similar to Eq.\eqref{EqForward0} introducing the probability of being at $x$ at time $t$. Summing over $N$ we have that

\begin{equation}
p^F(F,x|t)=\sum_{N=0}^\infty\int_0^tQ_N(t')\varphi(t+F-t')P(x,t-t')dt',
\end{equation}
and performing the Laplace transform on $t$ we get
\begin{equation}
\hat{p}^F(F,x|s)=\frac{\mathcal{L}_s\left[\varphi(t+F)P(x,t)\right]}{1-\hat{\varphi}(s)},
\end{equation}
where we have used Eq.\eqref{EqNthReset}. Using Eq. \eqref{eq:Q-def}, the above equation can be inverted by Laplace to get
\begin{equation}
p^F(F,x|t)=\int_{0}^{t}Q(t-t')\varphi(t'+F)P(x,t')dt'.
\label{pf1}
\end{equation}

Now, one can recover the general propagator $\rho(x,t)$ in Eq.\eqref{eq:generalpropagator-def} by integrating over $F$. The PDF for $F$ is thus
\begin{equation}
    f(F|x,t)=\frac{p^F(F,x|t)}{\rho(x,t)}=\frac{\int_{0}^{t}Q(t-t')\varphi(t'+F)P(x,t')dt'}{\rho(x,t)}.
    \label{eq:condforward-final}
\end{equation}
As we have done for the backward, in the following we study the behaviour of this PDF for different types of reset time distributions.

\subsubsection{$\varphi(t)$ with finite first moment}

Let us first consider resets happening with a finite mean time. In Section \ref{SecProp} we have seen that the system reaches a stationary state under this condition. So, in the long $t$ limit,

\begin{equation}
    f(F|x,t)\simeq \frac{\int_{0}^{t}Q(t-t')\varphi(t'+F)P(x,t')dt'}{\rho_s(x)}
\end{equation}
and, applying the Laplace transform for $t$,
\begin{equation}
    \hat{f}(F|x,s)\simeq \frac{\hat{Q}(s)\mathcal{L}[\varphi(t'+F)P(x,t')]}{\rho_s(x)}.
\end{equation}
Now, recalling that $\hat{Q}(s)\sim 1/\langle t \rangle_\varphi s$ for small $s$ (large $t$), we have that
\begin{equation}
    f(F|x)=\lim_{t\rightarrow \infty} f(F|x,t)\simeq \frac{\int_{0}^{\infty}\varphi(t'+F)P(x,t')dt'}{\langle t\rangle_\varphi \rho_s(x)}.
    \label{cfPDF}
\end{equation}

If the resets are exponentially distributed as in Eq.\eqref{eq:exponential}, the conditioned forward time PDF approximation takes the same exponential form
\begin{equation}
    f(F|x)\simeq r e^{-rF},
    \label{eq:condforward_expo}
\end{equation}
which does not depend on the position of the walker $x$. This is due to the Markovianity of the resetting process. Regardless of the moment (or position) we consider, the time until the next reset is equally distributed. 

\begin{figure*}
    \centering
    \includegraphics[scale=0.5]{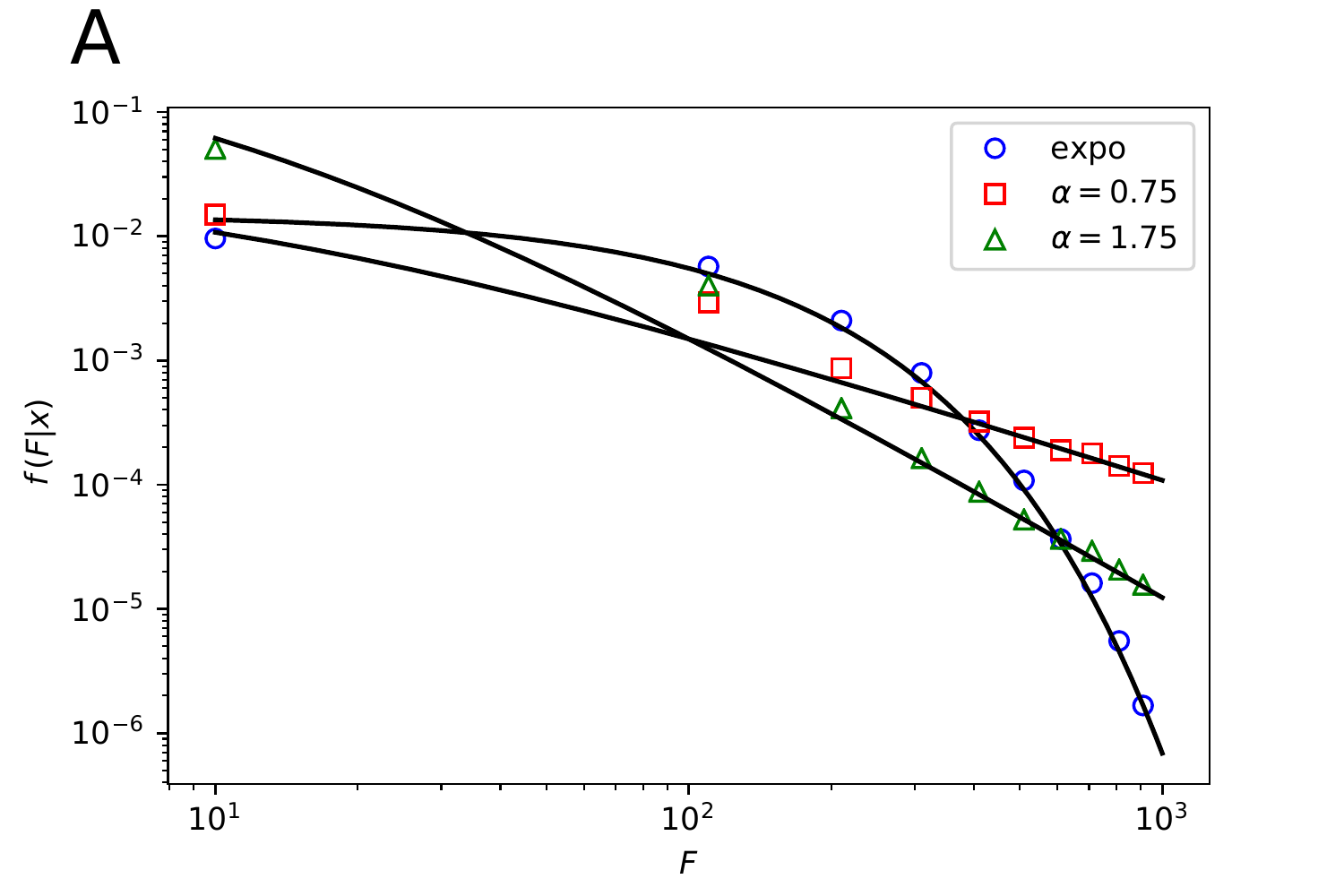}
    \includegraphics[scale=0.5]{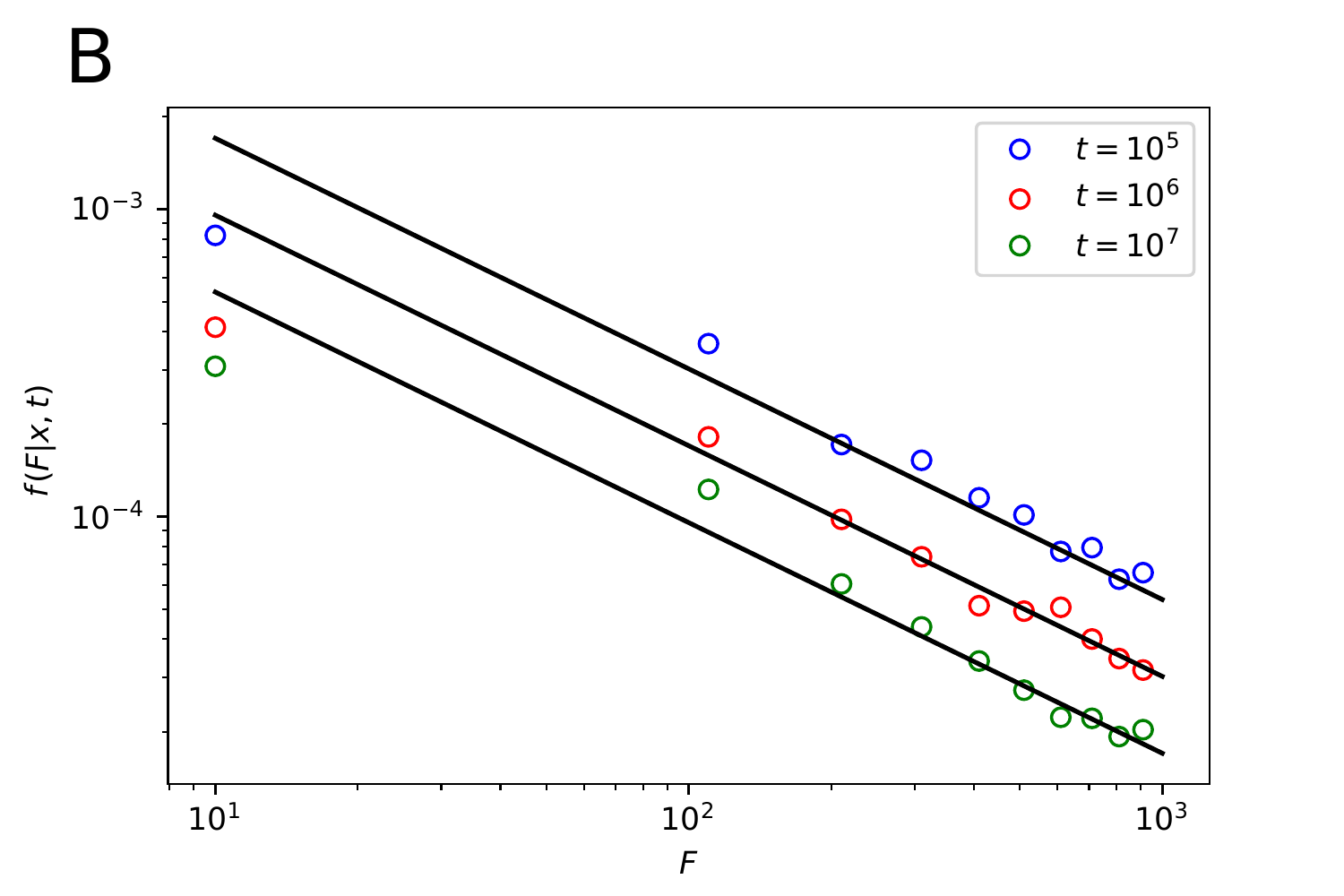}
    \caption{\textbf{A}: Stationary conditioned forward time distribution for diffusion under exponential resetting with $ r=0.05$ (circles) and under Pareto resetting with $ T=5$ and two different values of the decay exponent: $\alpha= 0.75$ (squares) and $\alpha=1.75$ (triangles). The solid lines have been drawn from the behaviours in Eq.\eqref{eq:condforward_expo} and Eq.\eqref{eq:condforward_pareto} for the exponential and Pareto cases respectively. 
    \textbf{B}: Conditioned forward time PDF for diffusion and Pareto resetting with $ T=5$ and $\alpha=0.25$. Three different measurement times have been plotted, showing that $f(B|x,t)$ does not reach a stationary shape. The solid lines correspond to the power-law behaviour in Eq.\eqref{eq:condforward_decay} for $\alpha < 1/2$. Both panels A and B have been simulated with diffusion constant $D=0.1$.}
    \label{fig:condforward}
\end{figure*}

If, instead, the reset times are drawn from a Pareto PDF according to Eq.\eqref{eq:pareto} with $\alpha >1$, the stationary conditioned forward time PDF can be approximated by 

\begin{equation}
    f(F|x)\simeq \frac{\alpha\left(\alpha-\frac{1}{2}\right)}{ T\left(1+\frac{F}{ T}\right)^{\frac{1}{2}+\alpha}}\frac{U\left(\alpha+\frac{1}{2},\frac{1}{2},\frac{x^{2}}{4D( T+F)}\right)}{U\left(\alpha-\frac{1}{2},\frac{1}{2},\frac{x^{2}}{4D T}\right)}.
    \label{eq:condforward_pareto}
\end{equation}
We refer the reader to Appendix \ref{AppendixB} for the detailed derivation. For small arguments, the Tricomi function tends to a constant value. Therefore, in the limit $ F \gg x^2/4D$, the only dependence on the forward time comes from the power-law factor and the conditioned forward time PDF scales with $F$ as  

\begin{equation}
    f(F|x)\sim \frac{ T^{\alpha-\frac{1}{2}}}{F^{\frac{1}{2}+\alpha}} \quad \textrm{as}\quad F \gg x^2/4D,
    \label{eq:condforward_decay}
\end{equation}
as can be seen in Fig. \ref{fig:condforward}A, where we show some examples of the distribution obtained from numerical simulations of the process. It is worth noting that the conditioned forward time PDF seems to have a long tail when the reset time PDF scales as $\varphi(t)\sim t^{-1-\alpha}$ with $\alpha > 1$. This happens for $\alpha < 3/2$. So, apparently, for $1<\alpha < 3/2$, the mean conditioned forward time is finite while the mean reset time diverges. This strangeness comes from the non-validity of the approximation when $F\propto t$. In Eq.\eqref{cfPDF}, when eliminating the current time by taking $t\rightarrow \infty$, we are implicitly considering the $F\ll t$ limit, where the scaling in Eq.\eqref{eq:condforward_decay} is valid. Nevertheless, this  behaviour varies when $F\propto t$, where the properties of the forward time are significantly different (see \cite{WaScDeBa18} for further details). Thus, the approximation herein employed to describe the conditioned forward time is only valid for $F\ll t$. This explains the apparent paradox of having a finite mean conditioned forward time when the mean reset time is finite.

\subsubsection{$\varphi(t)$ with all diverging moments}
Finally, we study the conditioned forward time PDF when all the moments of the reset time distribution diverge. Introducing Eq.\eqref{eq:prop_bulkdiverging} into Eq.\eqref{eq:condforward-final} and proceeding analogously we get that (see Appendix A for detailed calculations)
\begin{equation}
f(F|x,t)\simeq\frac{1}{ T}\frac{\sum_{n=0}^{\infty}\frac{\Gamma(n+\alpha+1)}{n!}\left(-\frac{t}{ T+F}\right)^{n}U\left(\alpha,\frac{1}{2}-n,\frac{x^{2}}{4Dt}\right)}{\left(1+\frac{F}{ T}\right)^{1+\alpha}\sum_{n=0}^{\infty}\frac{\Gamma(n+\alpha)}{n!}\left(-\frac{t}{ T}\right)^{n}U\left(\alpha,-n,\frac{x^{2}}{4Dt}\right)}.
\label{ffxt}
\end{equation}
 Similarly to what we have done for the propagator, we can get the long time limit $t\gg T$, $t\gg F$ and $ F \gg x^2/4D$ the scaling behaviour of the conditioned forward time PDF with $F$ in terms of $\alpha$ to be
\begin{equation}
f(F|x,t)\sim\left\{
\begin{matrix}
&\frac{t^{\alpha-\frac{1}{2}}}{F^{\frac{1}{2}+\alpha}},\ \text{for}\ 0<\alpha < \frac{1}{2} \\
 \\
&\frac{ T^{\alpha-\frac{1}{2}}}{F^{\frac{1}{2}+\alpha}},\ \text{for}\ \frac{1}{2}< \alpha < 1.
\end{matrix}
\right. 
\end{equation}
As in the finite-moment scenario, the conditioned forward time PDF attains a stationary shape when $\alpha > 1/2$. However, when the tail of the resetting distribution is wider ($\alpha < 1/2$), the conditioned forward time PDF depends explicitly on time even in the $t\rightarrow \infty$ limit, similarly to what happens with the conditioned backward time PDF. This has been checked numerically and the results are shown in Fig. \ref{fig:condforward}B. Once again, the value $\alpha = 1/2$ is relevant to describe both the conditioned forward and backward time PDFs.

      

\section{Conclusions}

\label{SecConclusions}

In this work we have introduced the conditioned forward and backward times for stochastic processes with resetting. Interestingly, for a diffusive process with resets, under certain conditions (see Section \ref{SecBackFor}) we are able to find a PDF for the forward and backward times which is independent of the measurement time $t$, depending only on the position of the walker. This result may be of particular relevance when considering processes for which the measurement time is inaccessible. In such cases, one can have statistical information about the forward and backward times by only knowing the current position of the walker. It may be interesting to study the conditioned backward and forward times for dynamics different than the diffusive random walker.

We have found that the behaviour of the conditioned backward and forward time PDFs for Pareto distributed reset times is different for $\alpha <1/2$ than when $\alpha > 1/2$. The appearance of $\alpha=1/2$ as a turning point is not new in the resetting literature \cite{MaCaMe19,BoChSo19} and it appears to be a general characteristic of diffusion with power-law resetting. Particularly, it arises when studying temporal features of the process. Somehow, the long time behaviour of diffusion $P(x,t)\sim 1/\sqrt{t}$ adds on the reset time PDF scaling $\varphi(t)\sim t^{-1-\alpha}$ when the focus is put on the time variable. This produces that, for diffusion with power-law resetting, significant changes on the dynamics occur when $\alpha$ is a half-integer instead of an integer. It would be interesting to study this aspect in much more detail to have more knowledge on the precise mechanism behind the junction of the temporal behaviour of diffusion and the resetting.

\section{Acknowledgments}
The authors would like to thank Eli Barkai for his comments and suggestions which have been of significant help. This research was partially supported by Grant No. CGL2016-78156-C2-2-R.
\appendix

\section{Propagator with infinite-mean resetting.}
\label{AppendixA}

\subsection{Bulk $x^2/4D\ll t$}

Computing the Fourier transform of Eq. \eqref{eq:prop_bulkdiverging} we get

\[
\tilde{\rho}(k,t)=\frac{a^{\alpha}e^{-Dk^{2}t}}{\Gamma(\alpha)\Gamma(1-\alpha)}\int_{0}^{1}\frac{e^{Dk^{2}ty}}{y^{1-\alpha}(1-ay)^{\alpha}}dy
\]
where we have introduced the new variable $y=1-B/t$ and have defined
\[
a=\frac{t/ T}{1+t/ T}.
\]
In the long time limit $t\gg T$ one has $a\simeq1$ and the integral
can be expressed in terms of the Kummer's M function
as
\begin{equation}
\tilde{\rho}(k,t)\simeq e^{-Dk^{2}t}M\left(\alpha,1,Dk^{2}t\right)\quad\textrm{if}\quad0<\alpha<1.\label{eq:tfr}
\end{equation}
Since we are interested in obtaining the expression of the propagator
in the bulk region we consider $x^{2}\ll Dt$ which is equivalent to $Dtk^{2}\gg\text{1.}$
The Kummer's M function $M(a,c,z)$ admits the asymptotic expansion
for large argument (see Eq. 13.1.4 in \cite{ab64})
\[
M(a,c,z)\simeq\frac{\Gamma(b)e^{z}z^{a-b}}{\Gamma(a)}\left[1+O(|z|^{-1})\right].
\]
Thus, from (\ref{eq:tfr}) 
\[
\tilde{\rho}(k,t)\simeq\frac{(Dt)^{\alpha-1}}{\Gamma(\alpha)|k|^{2(1-\alpha)}}
\]
which after inversion by Fourier yields
\begin{eqnarray}
& &\rho(x,t)\simeq\frac{1}{\pi\Gamma(\alpha)(Dt)^{1-\alpha}}\int_{0}^{\infty}\frac{\cos\left(kx\right)}{|k|^{2(1-\alpha)}}dk\nonumber\\
&=&\frac{\sin(\pi\alpha)\Gamma(2\alpha-1)}{\pi\Gamma(\alpha)(Dt)^{1-\alpha}}\frac{1}{|x|^{2\alpha-1}},\quad\textrm{if}\quad\frac{1}{2}<\alpha<1.
\label{eq:s1}
\end{eqnarray}
Alternatively, we can make use of the power series expansion of the
Kummer's M function (see Eq. 13.1.2 in \cite{ab64})
before inverting by Fourier. Hence, inserting
\[
M\left(\alpha,1,Dk^{2}t\right)=\frac{1}{\Gamma(\alpha)}\sum_{n=0}^{\infty}\frac{\Gamma(\alpha+n)}{(n!)^{2}}(Dk^{2}t)^{n}
\]
into Eq. (\ref{eq:tfr}) we find
\begin{eqnarray}
& &\rho(x,t)\simeq\frac{1}{\pi\Gamma(\alpha)}\sum_{n=0}^{\infty}\frac{\Gamma(\alpha+n)}{(n!)^{2}(Dt)^{-n}}\int_{0}^{\infty}k^{2n}\cos\left(kx\right)e^{-Dk^{2}t}dk\nonumber\\
&=&\frac{e^{-\frac{x^{2}}{4Dt}}}{2\pi\Gamma(\alpha)\sqrt{Dt}}\sum_{n=0}^{\infty}\frac{\Gamma(\alpha+n)\Gamma\left(n+\frac{1}{2}\right)}{(n!)^{2}}
M\left(-n,\frac{1}{2},\frac{x^{2}}{4Dt}\right).\label{eq:s2}
\end{eqnarray}
In the bulk region $x^{2}\ll 4Dt$ and then $e^{-\frac{x^{2}}{4Dt}}\simeq1+O(x^{2}/Dt)$
and $M\left(-n,\frac{1}{2},\frac{x^{2}}{4Dt}\right)\simeq1+O(x^{2}/Dt).$
On the other hand
\[
\sum_{n=0}^{\infty}\frac{\Gamma(\alpha+n)\Gamma\left(n+\frac{1}{2}\right)}{(n!)^{2}}=\frac{\Gamma\left(\frac{1}{2}-\alpha\right)\Gamma(\alpha)}{\Gamma(1-\alpha)}\quad\textrm{if}\quad0<\alpha<\frac{1}{2}.
\]
Finally, from this result and (\ref{eq:s2}) one readily finds
\begin{eqnarray}
\rho(x,t)\simeq\frac{\Gamma\left(\frac{1}{2}-\alpha\right)}{2\pi\Gamma(1-\alpha)}\frac{1}{\sqrt{Dt}}\quad\textrm{if}\quad0<\alpha<\frac{1}{2}.
\label{ss2}
\end{eqnarray}

\subsection{Tail $x^2\propto Dt$}

In order to derive the expression for the propagator when $x^2/4D\propto t$, we will demonstrate the 4 points enumerated in Appendix B from \cite{GoLu00} for the unconditioned backward time PDF. Here, we reproduce the derivation for the propagator $\rho(x,t)$.
\\
\\
i) \textit{Existence of a limiting distribution.}\\
To demonstrate that a limiting distribution exists, we study the asymptotic behaviour of the moments of the global propagator $\rho(x,t)$. To do so, we employ the well-known formula for the $n$-th moment in terms of the Fourier transform of the propagator
\begin{equation}
\langle x^n(s)\rangle =i^n \left[\frac{\partial^n \hat{\tilde{\rho}}(k,s)}{\partial k^n}\right]_{k=0}
\label{xns}
\end{equation}
in the Laplace space. The $n$-th derivative of \eqref{xns} can be expressed in terms of the Bell polynomials  by using the Fa\`a di Bruno's formula \cite{ch18}

\begin{equation}
\left[\frac{\partial^n \hat{\tilde{\rho}}(k,s)}{\partial k^n}\right]_{k=0}=\frac{\sum_{l=1}^n \hat{\varphi}^{*(l)}(s) B_{n,l}(0, 2D, 0,...,0)}{1-\hat{\varphi}(s)} 
\label{xn}
\end{equation}
where the exponent $(l)$ means derivative of order $l$. Noteworthy, the Bell polynomials $B_{n,l}(0,2D,0,...,0)\neq 0$ only when $n$ is even. Thus, the $(2n-1)-$th derivative of the propagator and, therefore, the $(2n-1)-$th moments are 0 as expected due to the symmetry of the process. Then for even $n$, only the term $l=n/2$ is different from 0. In particular $B_{n,n/2}(0,2D,0,...,0)=n!D^{n/2}/(n/2)!$. Therefore, from \eqref{xns} and \eqref{xn}
\begin{eqnarray}
\langle x^{2n}(s)\rangle&=&\frac{(-1)^{n}}{1-\hat{\varphi}(s)}\hat{\varphi}^{*(n)}(s)B_{2n,n}\left(0,2D,0,...,0\right)\nonumber \\
&=&(-1)^{n}D^{n}\frac{2n!}{n!}\frac{\hat{\varphi}^{*(n)}(s)}{s\hat{\varphi}^{*}(s)}
\end{eqnarray}
where we have used that $1-\hat{\varphi}(s)=s\hat{\varphi}^{*}(s)$. In the small $s$ limit (or long $t$) this can be inverted by Laplace to get
\begin{equation}
\langle x^{2n}(t)\rangle \approx (-1)^n \frac{\Gamma(\alpha)}{\Gamma(\alpha-n)\Gamma(n)}D^n t^n
\quad \text{as} \quad t\rightarrow\infty.
\end{equation}
In this limit all the even moments of the global propagator scale as $\langle x^{2n}(t)\rangle \sim t^n$. Therefore, there must exist a limiting distribution $\rho_Y(y)$ for the variable $y=x/\sqrt{t}$.
\\
\\
ii) \textit{Expression of the scaling function $g(\chi)$ in terms of the limiting distribution $\rho_Y(y)$.}\\
Let us find the integral expression of the scaling function in terms of the (yet unknown) limiting distribution of the afore-defined variable $y$. In the Fourier-Laplace space, the global propagator can be expressed as
\begin{equation}
\hat{\tilde{\rho}}(k,s)=\int_0^\infty dt e^{-st}\langle e^{-ikx}\rangle_X=\int_0^\infty dt e^{-st}\langle e^{-ik\sqrt{t}y}\rangle_Y,
\end{equation}
where, in the second equality, the expected value is computed with respect to the new variable $y$ instead of the original position variable $x$. Taking the expected value out of the integral and performing the Laplace transform within the brackets one gets
\begin{equation}
\hat{\tilde{\rho}}(k,s)= \frac{1}{s}-\frac{k\sqrt{\pi}}{2s^{3/2}}\left\langle ye^{-\frac{k^{2}y^{2}}{4s}}\left[i+\textrm{erfi}\left(\frac{ky}{2\sqrt{s}}\right)\right]\right\rangle _{Y}.
\end{equation}
The expected value has to be taken with the limiting distribution $\rho_Y(y)$. 

Now, in the long time limit (small $s$), the propagator can be described by the scaling function defined in Eq.\eqref{eq:gdeff}, as shown in Eq.\eqref{eq:propscaling}. From this relation, one can isolate the scaling function to be
\begin{equation}
g(\chi)= 1-\frac{\sqrt{\pi}}{2}\chi \left\langle ye^{-\frac{\chi^{2}y^{2}}{4}}\textrm{erfi}\left(\frac{\chi y}{2}\right)\right\rangle _{Y},
\label{eq:vmef}
\end{equation}
where we have used that $\langle y e^{-\chi^2 y^2/4}\rangle_Y=0$ due to symmetry. 
\\
\\
iii) \textit{Expression of the moments of the limiting distribution}.\\
Let us now expand the expressions of $g(\chi)$ from Eq.\eqref{eq:gdeff} and Eq.\eqref{eq:vmef}, and compare to get the moments of $\rho_Y(y)$. Starting from the first, its Taylor series for $\chi$ gives
\begin{equation}
g(\chi)=\sum_{n=0}^{\infty} (-1)^n\frac{\Gamma (n+1-\alpha)}{n!\Gamma (1-\alpha)} D^n \chi^{2n},
\end{equation}
while by expanding the derivative of the imaginary error function in the expected value of Eq.\eqref{eq:vmef} we get
\begin{equation}
g(\chi)=\sum_{n=0}^{\infty} (-1)^n \frac{\chi^{2n}}{2^n(2n-1)!!} \langle y^{2n}\rangle.
\label{g2}
\end{equation}

Now, comparing both expressions term by term, one can isolate the even moments of the limiting distribution to be
\begin{equation}
\langle y^{2n}\rangle =(4D)^n \frac{\Gamma\left(n+\frac{1}{2}\right)\Gamma (n+1-\alpha)}{\sqrt{\pi}\Gamma (1-\alpha)}.
\end{equation}
The odd moments are null due to the symmetry of the process. 
\\
\\
iv) \textit{Expression of the limiting distribution $\rho_Y(y)$}.

Finally, we can gather the information in the moments of $\rho_Y(y)$ to get an expression for it. The characteristic function of the limiting distribution can be then computed from the moments
\begin{equation}
\rho_{Y}(k)=\left\langle e^{iky}\right\rangle =\sum_{n=0}^{\infty}\frac{(-k^{2})^{n}}{(2n)!}\left\langle y^{2n}\right\rangle 
\end{equation}
so that

\begin{eqnarray*}
\rho_{Y}(k)&=&\frac{1}{\Gamma(1-\alpha)}\sum_{n=0}^{\infty}\frac{(-Dk^{2})^{n}}{n!}\Gamma\left(n+1-\alpha\right)\\
&=&M\left(1-\alpha,1,-Dk^{2}\right)
\end{eqnarray*}
 where $M(a,b,z)$ is the Kummer's M function. To invert by Fourier we express the Kummer's M function in integral form
 $$
 M(1-\alpha,1,-Dk^{2})=\frac{1}{\Gamma(\alpha)\Gamma(1-\alpha)}\int_{0}^{1}\frac{e^{-Dk^{2}u}}{u^{\alpha}(1-u)^{1-\alpha}}du.
 $$
 Then

\begin{eqnarray*}
\rho(y)&=&\frac{1}{\Gamma(\alpha)\Gamma(1-\alpha)}\frac{1}{\sqrt{4\pi D}}\int_{1}^{\infty}\frac{e^{-\frac{y^{2}}{4D}z}}{\sqrt{z}(z-1)^{1-\alpha}}dz\\
&=&\frac{1}{\Gamma(1-\alpha)}\frac{e^{-\frac{y^{2}}{4D}}}{\sqrt{4\pi D}}U\left(\alpha,\frac{1}{2}+\alpha,\frac{y^{2}}{4D}\right)
\end{eqnarray*}
where $U(a,b,z)$ is the Kummer's U function. If we undo the change of variable $y=x/ \sqrt{t}$ we get the PDF

\begin{equation}
    \rho(x,t)=\frac{1}{\Gamma(1-\alpha)}\frac{e^{-\frac{x^{2}}{4Dt}}}{\sqrt{4\pi Dt}}U\left(\alpha,\frac{1}{2}+\alpha,\frac{x^{2}}{4Dt}\right).
    \label{eq:finalprop_powerlaw_app}
\end{equation}

\section{Derivation of Eq. (\ref{eq:condforward_pareto})}
\label{AppendixB}

Inserting Eqs. \eqref{eq:diff_prop} and \eqref{eq:pareto} in the integral of Eq.\eqref{cfPDF} one has
\begin{eqnarray}
& &\int_{0}^{\infty}\varphi(t'+F)P(x,t')dt'=\frac{\alpha}{\sqrt{4\pi D T}}\frac{1}{\left(1+\frac{F}{ T}\right)^{\frac{1}{2}+\alpha}}\nonumber\\
&\times&\int_{0}^{\infty}u^{\alpha-\frac{1}{2}}(1+u)^{-1-\alpha}e^{-\frac{x^{2}}{4D( T+F)}u}du\nonumber\\
&=& \frac{\alpha}{\sqrt{4\pi D T}}\frac{\Gamma\left(\alpha+\frac{1}{2}\right)}{\left(1+\frac{F}{ T}\right)^{\frac{1}{2}+\alpha}}U\left(\alpha+\frac{1}{2},\frac{1}{2},\frac{x^{2}}{4D( T+F)}\right)
\label{inn}
\end{eqnarray}
where we have introduced the variable $u=( T+F)/t'$. Combining Eqs. \eqref{eq:ststate_powerlaw} and \eqref{inn} we finally obtain Eq. \eqref{eq:condforward_pareto}.

\section{Derivation of Eq. (\ref{ffxt})}
\label{AppendixG}
Let us begin with the calculation of the numerator in Eq. \eqref{eq:condforward-final}
in the limit $t\gg T$. Introducing Eqs. \eqref{eq:diff_prop}, \eqref{eq:pareto}
and \eqref{eq:Q-diverging} into the integral of Eq.\eqref{eq:condforward-final}
we get

\begin{eqnarray}
& &\int_{0}^{t}Q(t-t')\varphi(t'+F)P(x,t')dt'\simeq\frac{\alpha}{t^{3/2}\sqrt{4\pi D}\Gamma(\alpha)\Gamma(1-\alpha)}\nonumber\\
&\times&\int_{1}^{\infty}\frac{e^{-\frac{x^{2}}{4Dt}y}\left(\frac{ T+F}{t}+y^{-1}\right)^{-1-\alpha}}{y^{\frac{1}{2}+\alpha}(y-1)^{1-\alpha}}dy,\label{eq:nupf}
\end{eqnarray}
where we have defined the variable $y=t/t'.$ It is useful to write the factor as
\begin{eqnarray}
\left(\frac{ T+F}{t}+y^{-1}\right)^{-1-\alpha}=\sum_{n=0}^{\infty}\lambda_{n}y^{-n}\label{aux}
\end{eqnarray}
as power series of $y^{-1}$ where $\lambda_{n}$ are the corresponding
coefficients of the Maclaurin expansion: 
\[
\lambda_{n}=\left(\frac{t}{F+ T}\right)^{1+\alpha}\frac{\Gamma(1+n+\alpha)}{\Gamma(1+\alpha)n!}\left(-\frac{t}{F+ T}\right)^{n}.
\]
 Plugging the above expansion into Eq. (\ref{eq:nupf}) express the
integral in the following form
\begin{eqnarray}
& &\int_{0}^{t}Q(t-t')\varphi(t'+F)P(x,t')dt'\nonumber\\
&\simeq&\frac{\alpha e^{-\frac{x^{2}}{4Dt}}}{t^{3/2}\sqrt{4\pi D}\Gamma(1-\alpha)}\sum_{n=0}^{\infty}\lambda_{n}U\left(\alpha,\frac{1}{2}-n,\frac{x^{2}}{4Dt}\right)
\end{eqnarray}
where we have made use of Eq. 13.2.6 in \cite{ab64}. The denominator
in Eq. \eqref{eq:condforward-final} is nothing but the propagator. It
can be obtained by inserting Eqs. \eqref{eq:diff_prop}, \eqref{eq:pareto_survival}
and \eqref{eq:Q-diverging} into \eqref{eq:generalpropagator-def}.
After defining the new variable $y=t/B$ one has 
\begin{eqnarray}
\rho(x,t)&\simeq&\frac{1}{ T^{\alpha}t^{1/2-\alpha}\sqrt{4\pi D}\Gamma(\alpha)\Gamma(1-\alpha)}\nonumber\\
&\times&\int_{1}^{\infty}\frac{e^{-\frac{x^{2}}{4Dt}y}\left(\frac{ T}{t}+y^{-1}\right)^{-\alpha}}{y^{1+\alpha}(y-1)^{1-\alpha}}dy\label{eq:propgf}
\end{eqnarray}
which holds in the limit $t\gg T$. Making use of the Maclaurin
expansion the factor $\left(\frac{ T}{t}+y^{-1}\right)^{-\alpha}$ reads
\[
\left(\frac{ T}{t}+y^{-1}\right)^{-\alpha}=\frac{1}{\Gamma(\alpha)}\sum_{n=0}^{\infty}\frac{\Gamma(n+\alpha)}{n!}\left(-\frac{t}{ T}\right)^{n}y^{-n},
\]
and the integral in Eq. (\ref{eq:propgf}) can be computed using again Eq.
13.2.6 in \cite{ab64}. One readily finds
\begin{eqnarray}
\rho(x,t)&\simeq&\frac{e^{-\frac{x^{2}}{4Dt}}}{ T^{\alpha}t^{1/2-\alpha}\sqrt{4\pi D}\Gamma(\alpha)\Gamma(1-\alpha)}\nonumber\\
&\times&\sum_{n=0}^{\infty}\frac{\Gamma(n+\alpha)}{n!}\left(-\frac{t}{ T}\right)^{n}U\left(\alpha,-n,\frac{x^{2}}{4Dt}\right).\label{eq:propgf2}
\end{eqnarray}
Dividing Eqs. (\ref{eq:nupf}) and (\ref{eq:propgf2}) we finally find Eq. \eqref{ffxt} in the main text.

\bibliography{main}

\end{document}